\documentclass[twocolumn,aps,prl,showpacs,amsmath,amssymb,superscriptaddress]{revtex4-1}

\usepackage{physics}
\usepackage{graphicx} 
\usepackage{dcolumn}
\usepackage[caption=false]{subfig}
\usepackage[colorlinks=true,linkcolor = {blue}, citecolor = {blue}, urlcolor = {blue}]{hyperref}
\usepackage[dvipsnames]{xcolor}
\usepackage[normalem]{ulem}
\begin{document}

\title{Emergent inflation of the Efimov spectrum under three-body spin-exchange interactions}

\author{J. van de Kraats}
\email[Corresponding author:  ]{j.v.d.kraats@tue.nl}
\affiliation{Eindhoven University of Technology, P. O. Box 513, 5600 MB Eindhoven, The Netherlands}
\author{D. J. M. Ahmed-Braun}
\affiliation{Eindhoven University of Technology, P. O. Box 513, 5600 MB Eindhoven, The Netherlands}
\author{J.-L. Li}
\affiliation{Institut f{\"u}r Quantenmaterie and Center for Integrated Quantum Science and Technology IQ ST, Universit{\"a}t Ulm, D-89069 Ulm, Germany}
\author{S. J. J. M. F. Kokkelmans}
\affiliation{Eindhoven University of Technology, P. O. Box 513, 5600 MB Eindhoven, The Netherlands}
\date{\today}

\begin{abstract}
We resolve the unexpected and long-standing disagreement between experiment and theory in the Efimovian three-body spectrum of \textsuperscript{7}Li, commonly referred to as the lithium few-body puzzle. Our results show that the discrepancy arises out of the presence of strong non-universal three-body spin-exchange interactions, which enact an effective inflation of the universal Efimov spectrum. This conclusion is obtained from a thorough numerical solution of the quantum mechanical three-body problem, including precise interatomic interactions and all spin degrees of freedom for three alkali-metal atoms. Our results show excellent agreement with the experimental data regarding both the Efimov spectrum and the absolute rate constants of three-body recombination, and in addition reveal a general product propensity for such triatomic reactions in the Paschen-Back regime, stemming from Wigner’s spin conservation rule.
\end{abstract}

\maketitle

\textit{Introduction.}--- There exists a general desire in physics to formulate accurate descriptions of nature from a minimal number of adjustable parameters, thus uncovering the presence of \textit{universal} behavior. A paradigmatic example of a system where this ideal picture is realized is the scattering of two particles at low energy. Here, the wave function delocalizes to the point that the observable properties of the system become insensitive to the exact microscopic detail of the interaction, allowing for a description purely in terms of the $s$-wave scattering length $a$ \cite{Braaten2006}. This remarkable universality carries through to systems of more than two particles, most strikingly exemplified at the three-body level by virtue of the Efimov effect. At a two-body scattering resonance, where $a \rightarrow \infty$, the Efimov effect induces a universal emergence of an infinite tower of geometrically spaced three-body bound states \cite{Efimov1970, Efimov1971}. The resulting spectrum is fully determined by a single length scale, the three-body parameter, typically expressed as the negative scattering length $a_-$ where the ground state trimer dissociates into the three-body scattering continuum \cite{Braaten2006, Naidon2017, Greene2017, Incao2018}. In turn, the Efimov effect and three-body parameter induce universal properties in few-body clusters of four or more particles, further extending the applicability of universal theory \cite{Greene2017}.

The vast majority of experimental studies of the Efimov effect utilize ultracold atomic gases, where the scattering length can be directly controlled by means of a magnetic Feshbach resonance \cite{Kohler2006, Chin2010}. Near such a resonance, the three-body parameter can be extracted from a characteristic log-periodic modulation of the rate of three-body recombination \cite{Esry1999, Braaten2006, Naidon2017}. Interestingly, although the precise value of $a_-$ is typically sensitive to non-universal short-range physics, the Efimov spectrum in atomic systems possesses an additional van der Waals universality $a_- \approx -9.7 \ r_{\mathrm{vdW}}$ \cite{Kraemer2006, Berninger2011,  Wild2012}, where $r_{\mathrm{vdW}}$ gives the characteristic length scale associated with the two-body interaction. Theoretical analyses have shown that this universality is robust for Efimov states near broad Feshbach resonances, and originates from a universal suppression of three-body probability in the short range \cite{Wang2012, Naidon2014}. 

For Feshbach resonances of intermediate to narrow width, both theoretical and experimental works have demonstrated an increase of $\abs{a_-}$, arising from the associated growth of the two-body effective range scale \cite{Petrov2004, Gogolin2008, Schmidt2012, Langmack2018, Chapurin2019, Secker2021_2C, Etrych2023, Kraats2023}. However, a series of experiments in this regime with the lightest bosonic alkali, \textsuperscript{7}Li, have failed to observe this behavior, and in fact measured values for $\abs*{a_-}$ that, remarkably, recede slightly below the universal van der Waals value \cite{Pollack2009, Gross2009, Gross2010, Gross2011, Dyke2013}. While similar behavior can be obtained in some theoretical scenarios \cite{Sorensen2012, Secker2021_2C}, it is generally unclear how to connect these to \textsuperscript{7}Li, and sophisticated numerical models have so far failed to reproduce the data. The long-standing challenge to explain this unexpected mismatch between theory and experiment is now referred to as the \textit{lithium few-body puzzle} \cite{Yudkin2021, Incao2022, Yudkin2023_abc}. As descriptions of quantum matter are typically derived starting from an understanding of the underlying microscopic physics, elucidating the origin of the lithium few-body puzzle is relevant to a wide array of different subfields. In particular, resolving this puzzle opens up opportunities to further study fundamental questions such as the possible impact of non-additive three-body forces \cite{Incao2022}, non-universal multichannel physics \cite{Yudkin2023_abc}, and the prevalence of few-body universality \cite{Yudkin2023_resh}.

\begin{figure}[t]
\includegraphics[width=0.8\linewidth]{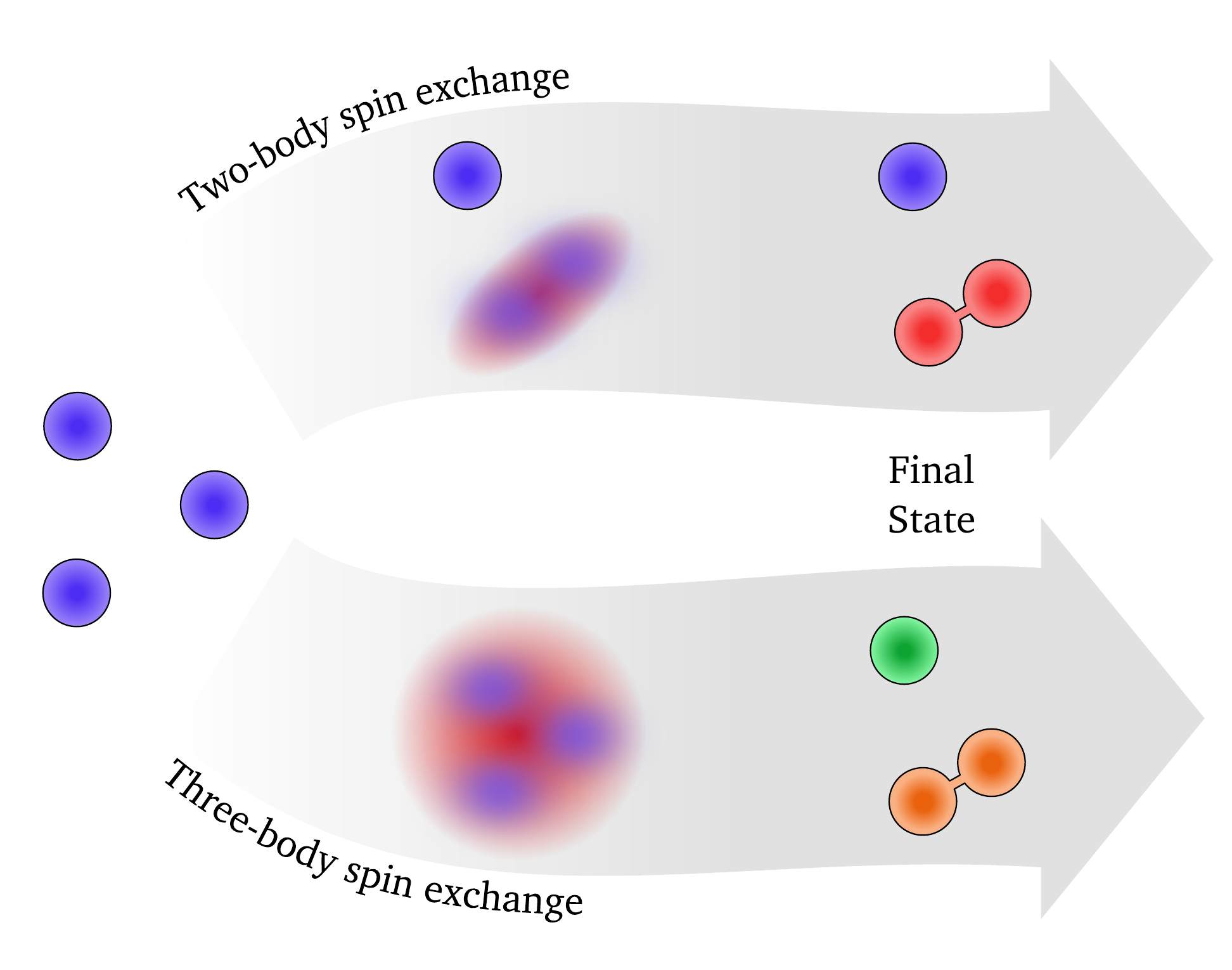}
\caption{\label{fig:FSSvsFMS}Illustration of three-body recombination through two distinct spin-exchange pathways, where the color of particles represents their spin state and two connected particles represent a molecule. In the upper pathway one of the particles conserves its spin throughout the recombination. In the bottom pathway, all three particles partake in spin-exchange such that no single spin is conserved. The pathway in which all particles preserve their initial spin is included in all calculations.}
\end{figure}

In this Letter, we investigate the connection between the anomalous value of $\abs*{a_-}$ in \textsuperscript{7}Li and the presence of three-body spin-exchange interactions. Here, \textit{spin-exchange} refers to a process in which the internal spin state of the atoms is altered by coupling of the valence electrons, and thus necessarily occurs at short length scales. Due to the aforementioned suppression of three-body probability in this regime, a useful distinction can be made between two-body spin-exchange, where the state of the third particle is conserved, and three-body spin-exchange, where all particles change their spin. As we illustrate in Fig.~\ref{fig:FSSvsFMS}, this distinction can also naturally be applied to three-body recombination. For many applications, the contributions of three-body spin-exchange are negligible, which significantly simplifies the three-body problem \cite{Petrov2004, Gogolin2008, Massignan2008, Schmidt2012, Wolf2017, Langmack2018, Chapurin2019, Haze2022, Secker2021_2C}. Recently however, studies have found three-body spin-exchange to contribute significantly to three-body observables in \textsuperscript{39}K \cite{Secker2021_39K} and \textsuperscript{7}Li \cite{Li2022}, both at relatively large magnetic fields. Motivated by these findings we study the Efimov spectrum in \textsuperscript{7}Li, using a recently developed numerical approach to the quantum mechanical three-body problem which, together with high-performance computing facilities, allows us to include \textit{all} coupled three-body channels in the Hamiltonian \cite{Secker2021_Grid, Secker2021_39K}.

\textit{Inflation of the Efimov spectrum.}--- Following the experiment of Ref.~\cite{Gross2011}, we analyze two high-field Feshbach resonances in spin-polarized ultracold \textsuperscript{7}Li gases, in the hyperfine states $\ket*{f, m_f}_{\mathrm{in}} = \ket*{1,1}$ and $\ket*{f, m_f}_{\mathrm{in}} = \ket*{1,0}$ respectively. Both resonances have similar negative background scattering lengths on the order of $r_{\mathrm{vdW}} = 32.4863 \ a_0$, and are both of intermediate to narrow resonance width \cite{Gross2011, Julienne2014}. In line with experiment we study the rate of three-body loss of the trapped gas density $n$, typically expressed in terms of a recombination rate constant $L_3$ as $\mathrm{d}n/\mathrm{d}t = - L_3 \ n^3$. In the zero-energy limit it can be formally expressed as \cite{Lee2007, Moerdijk1996, Smirne2007, Secker2021_Grid},
\begin{equation}
L_3  =  \sum_{{\nu},  c_3} L_3^{\mathrm{part}}(\varphi_\nu, c_3),
\label{eq:K3_1}
\end{equation}

where,
\begin{eqnarray}
&&L_3^{\mathrm{part}} (\varphi_{\nu}, c_3)   =   \frac{12 \pi m}{\hbar} (2 \pi \hbar)^6 \label{eq:K3_2} \\* && \times \lim_{E \rightarrow 0}   \int d\vu{q}_3 \ q_3 \abs*{\matrixel*{(\vb{q}_3, c_3),  \varphi_{\nu}}{U_{\alpha0}(E)}{\Psi_{\mathrm{in}}}}^2, \nonumber
\end{eqnarray}
reminiscent of Fermi's golden rule. Each element $\matrixel*{(\vb{q}_3, c_3), \varphi_{\nu}}{U_{\alpha0}(z)}{\Psi_{\mathrm{in}}}$ describes transition from an incoming three-body state with energy $E$ into a molecular state $\ket*{\varphi_{\nu}}$ of a pair $\alpha$ and a third free particle with relative momentum $\vb{q}_3$ in spin state $\ket*{c_3}$. 

To calculate the operator $U_{\alpha0}(E)$ we solve the inelastic three-body scattering problem in momentum space, using the Alt-Grassberger-Sandhas (AGS) equations \cite{Alt1967} (see Supplemental Material \cite{SupMat}). As any two-body subsystem can transfer energy to the third particle, the three-body problem embeds the \textit{off-shell} solution to the two-body problem, which we obtain by exact diagonalization of the two-body Hamiltonian. We calculate $L_3$ for a range of scattering lengths on the attractive ($a<0$) side of the Feshbach resonance, and subsequently extract the values of the three-body parameter $a_-$ and trimer width $\eta_-$ by fitting the data to universal predictions from effective field theory \cite{Braaten2006}. To highlight the role of three-body spin-exchange, we compare two different approaches to the spin-basis of the three-body problem, referred to as Full Multichannel Spin (FMS) and Fixed Spectating Spin (FSS) models \cite{Secker2021_39K}. In an FMS calculation, all coupled three-body spin-channels that conserve the total magnetic quantum number $M_F = m_{f_1} + m_{f_2} + m_{f_3}$ are included, thus taking into account both recombination pathways in Fig.~\ref{fig:FSSvsFMS}. In an FSS model, the spin basis is constrained to only those channels where the spin of the third particle is fixed, thus omitting the lower pathway.
To model the pairwise interactions we use realistic singlet and triplet Born-Oppenheimer interaction potentials \cite{Julienne2014}. It follows that the FMS model is quantum mechanically rigorous, save for the omission of non-additive three-body forces which are generally assumed to be negligible in ultracold atomic gases \cite{Naidon2017}. 

\begin{figure*}[t]
\includegraphics{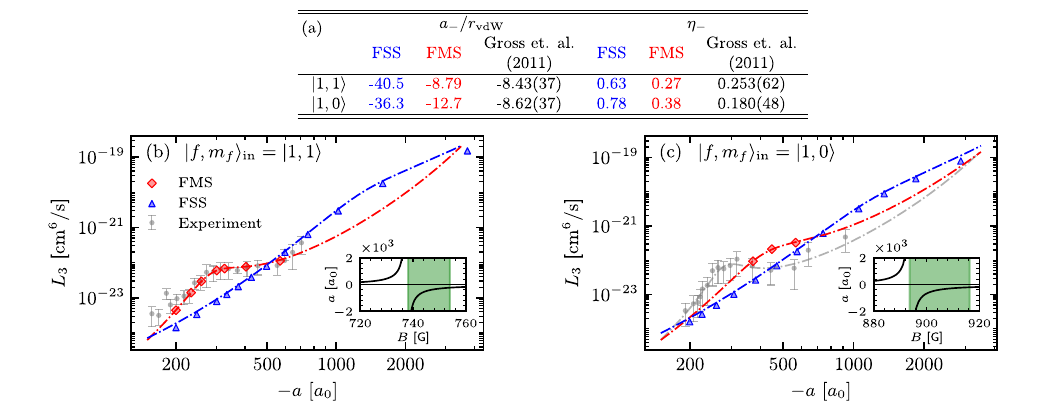}
\caption{Results of our calculations, compared directly with the experimental data of Ref.~\cite{Gross2011}. Table (a) shows the three-body parameter $a_-$ and trimer width $\eta_-$ obtained using FSS and FMS models. The associated values of the three-body recombination rate coefficient $L_3$ as a function of the scattering length $a$ are shown in Figs. (b) and (c), for the two distinct incoming hyperfine states. The fits of $L_3$ to universal theory giving the results in table (a) are shown as dash-dotted lines in matching color. Insets show the two-body scattering length as a function of magnetic field, where the green shaded region matches the range of scattering lengths in the enclosing figure. We note that an additional independent measurement of the Efimov trimer in the $\ket*{1,1}$ state obtained $a_-/r_{\mathrm{vdW}} = -7.76(31)$ and $\eta_- = 0.17$ \cite{Dyke2013}.}
\label{fig:3BP}
\end{figure*}

Our results are shown in Fig.~\ref{fig:3BP}. In an FSS calculation, the value of $\abs*{a_-}$ is significantly larger than both the universal van der Waals value and the experimental data, suggesting a significant squeezing of the spectrum for this narrow resonance which is in line with the majority of multichannel three-body models in the current literature \cite{Schmidt2012, Langmack2018, Kraats2023}. Our main result is that upon including three-body spin-exchange processes, the additional accessible states induce non-trivial multichannel physics that acts to cancel the increase of $\abs*{a_-}$, and can even decrease $\abs*{a_-}$ to below the universal van der Waals value thus resulting in an inflation of the spectrum. Consequently, our FMS calculations significantly improve on the FSS results with respect to the experimental data, and for the $\ket*{f, m_f}_{\mathrm{in}} = \ket*{1,1}$ state specifically our FMS results for both $a_-$ and $\eta_-$ fall within the experimental uncertainty of Ref.~\cite{Gross2011}. As conservation of $M_F$ dictates that the number of coupled three-body channels in the $\ket*{f, m_f}_{\mathrm{in}} = \ket*{1,0}$ state is about twice as large as in the $\ket*{f, m_f}_{\mathrm{in}} = \ket*{1,1}$ state, the corresponding FMS calculation is considerably more expensive numerically. For this reason we can not fully converge this calculation in the ultraviolet cut-off on the momentum grid and the number of angular momentum states, which leads to a discrepancy between our calculations and the experimental data in Fig.~\ref{fig:3BP}(c). However, as we show in the Supplementary Material \cite{SupMat}, the change in $a_-$ with the relevant numeric parameters indicates that a match with experiment is also achievable for the $\ket*{f, m_f}_{\mathrm{in}} = \ket*{1,0}$ state if numerical resources allow. Regardless, the significant improvement of the FMS model compared to the FSS model is clear.

Previous three-body studies have shown that the universal increase in $\abs{a_-}$ near narrow resonances arises from a repulsive barrier in the three-body potential scaling with the effective range, which progressively squeezes the Efimov spectrum \cite{Tempest2023}. Recently however, the unexpected observation of a trimer state above the atom-dimer disassociation threshold has prompted further theoretical analysis, which indicates that the Efimov state of $^{7}$Li may actually exist \textit{behind} the universal repulsive barrier \cite{Yudkin2023_resh}. While the universal effects of the barrier are evident in our FSS results, the observed sensitivity of the value of $\abs{a_-}$ to short-range three-body spin-exchange processes is in fact consistent with the presence of a non-universal trimer state in the inner potential well. In this sense, our FMS results can serve as important numerical confirmation of this novel trimer binding mechanism. Such an identification furthermore indicates that three-body spin-exchange couplings induce an effective attractive interaction that can tug the trimer state into the inner potential well, thus causing the inflation of the Efimov spectrum. In the future, it may be interesting to analyze the exact nature of this trimer and the corresponding potential in more detail, for which an approach similar to Ref.~\cite{Kraats2023} could prove a useful starting point.

\begin{figure*}[t]
\includegraphics{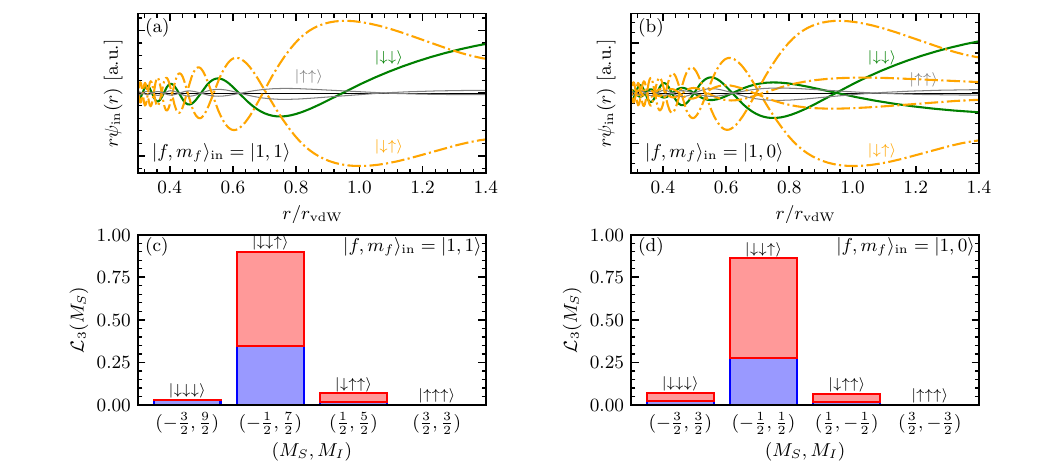}
\caption{Analysis of the electronic spin-propensity rule for three-body recombination. In Figs.~(a) and (b) we show the components of the incoming two-body radial wave functions $r \psi_{\mathrm{in}}(r)$, highlighting the electronic spin components $\ket*{{\downarrow}{\downarrow}}$ in green and $\ket*{{\downarrow}{\uparrow}}$ in dash-dotted orange. The resonantly-enhanced singlet $\ket*{{\downarrow}{\uparrow}}$ components dominate the wave function at short distance. In Fig. (c) and (d) we show the averaged partial recombination rate $\mathcal{L}_3(M_S)$ to channels with total electronic projection $M_S$, with associated nuclear projection $M_I$. The fraction in blue(red) originates from atom-dimer states accessible by two-body(three-body) spin-exchange.}
\label{fig:SpinPropensity}
\end{figure*}

\textit{Propensity rule for three-body recombination.}--- Next to the excellent match with the three-body parameter our calculations also show good agreement with the individual measurements of $L_3$. As three-body recombination is an important and ubiquitous chemical process, relevant far beyond the specific context of Efimov physics, this agreement motivates us to analyze the recombination rates more closely \cite{Mirahmadi2022}. To characterize the nature of the spin-exchange pathways we separately examine the partial recombination rates $L_3^{\mathrm{part}}$, which effectively provide a measure of the population distribution of product states following three-body recombination. 

Interestingly, our calculations hint at the existence of a spin propensity rule, as the number of strongly coupled channels is remarkably small, with the vast majority of product channels having near negligible relative recombination rates (see Supplementary Material \cite{SupMat} for more detail). We will now show that this behavior results from a manifestation of Wigner's electronic-spin conservation rule \cite{Wigner1927,Moore1973} for three atoms, originating from the relatively weak coupling between the electronic and nuclear spins of the atoms at large magnetic fields. In this Paschen-Back regime \cite{Paschen1921}, the spins independently precess around the magnetic field direction, such that single-particle states are best described by the individual projection quantum numbers $m_s$ and $m_i$. For simplicity let us briefly neglect the subdominant contribution from the nuclear spin, whose coupling to the magnetic field is relatively weak. Then both incoming single-particle states studied in this work may be written as, 
\begin{equation}
\ket*{f, m_f}_{\mathrm{in}} \sim \ket*{\downarrow} + \delta \ket*{\uparrow},
\end{equation}
where $\ket*{{\downarrow}/{\uparrow}}$ represent the down and up electronic spin states $m_s = -1/2$ and $m_s = 1/2$ respectively, and $\delta$ scales with the ratio of the hyperfine and Zeeman energies \cite{Stoof1988}. 

In the Paschen-Back regime $\delta$ is a small number, which motivates us to expand the incoming three-body state into four distinct components scaling as $\delta^{n}$, where $n$ equals the number of electronic spins pointing up. Each component can be uniquely identified with a definite value of the total electronic spin projection $M_S = m_{s_1} + m_{s_2} + m_{s_3} = -3/2 + n$, which is rigorously conserved in this basis as it is fully uncoupled from the nuclear spin. Hence, if it is possible to identify a dominant incoming projection $M_S$, then the outgoing product state distribution will show a propensity to states that conserve this projection. To determine the dominant incoming component we have to consider both the scaling with the small parameter $\delta$ and the amplitude of the associated wave functions at small nuclear separations, where recombination processes typically take place. As we illustrate in Figs.~\ref{fig:SpinPropensity}(a) and \ref{fig:SpinPropensity}(b), the dominant short-range components of the incoming two-body wave functions are in the singlet two-body state $\ket*{{\downarrow} {\uparrow}}$, correspondent with the spin character of the resonantly coupled Feshbach level. It follows that recombination preferably occurs through three-body states with (partial) singlet character, of which the state $\ket*{{\downarrow} {\downarrow} {\uparrow}}$ ($n = 1$) has the dominant scaling with $\delta$. Thus, we finally deduce that three-body recombination will show a propensity to product channels with $M_S = -\frac{1}{2}$.

To confirm the presence of this propensity in our numerics we define an augmented partial recombination rate $\mathcal{L}_3(\left\{m_{s_j}, m_{i_j} \right\})$ as,
\begin{eqnarray}
\mathcal{L}_3(\left\{m_{s_j}, m_{i_j} \right\}) &=& \sum_{\nu, c_3} \abs*{\braket*{m_{s_1}, m_{i_1}, m_{s_2}, m_{i_2}}{\varphi_{\nu}}}^2 \\ &&  \times \abs*{\braket*{m_{s_3}, m_{i_3}}{c_3}}^2   \frac{L_3^{\mathrm{part}}(\varphi_{\nu}, c_3)}{L_3}. \nonumber
\end{eqnarray}
By this definition $\mathcal{L}_3(\left\{m_{s_j}, m_{i_j} \right\})$ averages the spin-projection $\abs*{\braket*{m_{s_1}, m_{i_1}, m_{s_2}, m_{i_2}}{\varphi_{\nu}}}^2 \abs*{\braket*{m_{s_3}, m_{i_3}}{c_3}}^2$ with respect to the discrete probability distribution $L_3^{\mathrm{part}}(\varphi_{\nu}, c_3)/L_3$, and hence forms a measure of the relative importance of a state defined by a set of quantum numbers $\left\{m_{s_j}, m_{i_j} \right\}$ for three-body recombination, normalised to unity if summed over all available states. To obtain $\mathcal{L}_3(M_S)$ we sum $\mathcal{L}_3(\left\{m_{s_j}, m_{i_j} \right\})$ over all spin states with definite $M_S$, which then gives the results shown in Figs.~\ref{fig:SpinPropensity}(c) and \ref{fig:SpinPropensity}(d). The electronic spin-conservation propensity is clearly present for both Feshbach resonances we study. Note that in the argument outlined above all three particles are treated equally, and no reference is made to a special role for the spectating particle. Indeed, we observe that the $\ket*{{\downarrow} {\downarrow} {\uparrow}}$ component of $\mathcal{L}_3$ is split almost equally between channels in which the spectating spin is conserved or changed, with the latter being slightly larger.

\textit{Outlook.}--- Our findings suggest several new avenues for future research. First the excellent match between our \textsuperscript{7}Li results and the experimental data now provides a new benchmark for the theoretical description of strongly interacting few-body systems. Our method thus shows great promise for studying other systems where measurements deviate from the current theoretical predictions \cite{Etrych2023}. Aside from these experimental concerns, there is also a more conceptual challenge to now further characterize the physical mechanism underpinning the formation of the non-universal Efimov trimer observed in this work, which will require untangling the exact reshaping of the three-body potential in the presence of three-body spin-exchange \cite{Kraats2023, Tempest2023, Yudkin2023_resh}. 

Our results also have interesting implications beyond the realm of Efimov physics. The uncovered spin-propensity rule in the rate of three-body recombination provides a remarkably simple picture of triatomic chemical reactions in large magnetic fields, which can now aid in the understanding and possible experimental control of state-to-state quantum chemistry in these regimes \cite{Wolf2017, Haze2022}. Further studies in this direction may also seek to elucidate the more subtle role of the individual nuclear spins, which should have a similar propensity to be conserved in the Paschen-Back regime.

\begin{acknowledgments}
We thank Jose D'Incao for fruitful discussions. J.v.d.K. and S.J.J.M.F.K. acknowledge financial support from the Dutch Ministry of Economic Affairs and Climate Policy (EZK), as part of the Quantum Delta NL program. D.J.M.A.B. acknowledges financial support from the Netherlands Organisation for Scientific Research (NWO) under Grant No. 680-47-623.  The results presented in this work were obtained on the Dutch national supercomputer Snellius, with support from TU/e HPC Lab and Surf. 
\end{acknowledgments}

\bibliography{References}

\end{document}


\title{Supplementary Material: ``Emergent inflation of the Efimov spectrum under three-body spin-exchange interactions"}

\author{J. van de Kraats}
\email[Corresponding author:  ]{j.v.d.kraats@tue.nl}
\affiliation{Eindhoven University of Technology, P. O. Box 513, 5600 MB Eindhoven, The Netherlands}
\author{D. J. M. Ahmed-Braun}
\affiliation{Eindhoven University of Technology, P. O. Box 513, 5600 MB Eindhoven, The Netherlands}
\author{J.-L. Li}
\affiliation{Institut f{\"u}r Quantenmaterie and Center for Integrated Quantum Science and Technology IQ ST, Universit{\"a}t Ulm, D-89069 Ulm, Germany}
\author{S. J. J. M. F. Kokkelmans}
\affiliation{Eindhoven University of Technology, P. O. Box 513, 5600 MB Eindhoven, The Netherlands}
\date{\today}

\begin{abstract}
\end{abstract}
\maketitle

\section{Three-body model and numerical method}

In this Supplementary Material we give a more detailed overview of our model and numerical method, outlining the Hamiltonian, three-body integral equations and spin models. We also give more detail on the universal fit of $L_3$, and discuss the numerical convergence. We note that the method as outlined here is based on Ref.~\cite{Secker2021_Grid}.

\subsection{Three-body Hamiltonian}

\begin{figure*}[t]
\includegraphics{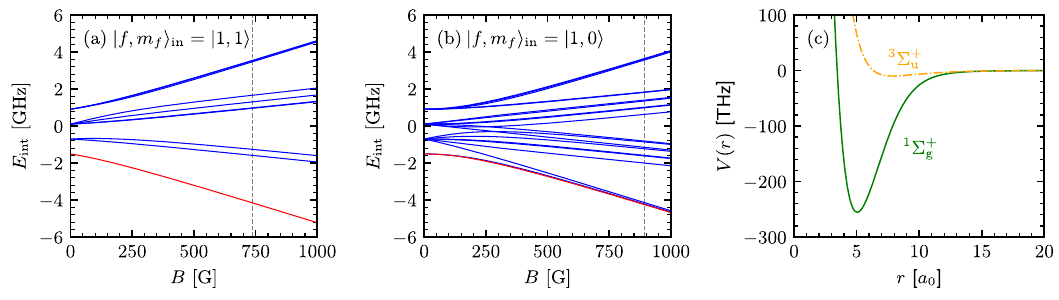}
\caption{\label{fig:3bodyMC} Multichannel aspects of the three-body problem. In Figs (a) and (b) we show the spectrum of coupled non-degenerate three-body internal states as a function of the external magnetic field $B$, for the $\ket*{f, m_f}_{\mathrm{in}} = \ket*{1, 1}$ and $\ket*{f, m_f}_{\mathrm{in}} = \ket*{1, 0}$ incoming states respectively. The state highlighted in red represents the incoming state and thus also gives the three-body scattering threshold. The vertical dash-dotted line indicates the location of the Feshbach resonance studied in the main text. In (c) we show the molecular singlet ($^{1}\Sigma_{\mathrm{g}}^{+}$) and triplet ($^{3}\Sigma_{\mathrm{u}}^{+}$) Born-Oppenheimer interaction potentials as formulated in Ref.~\cite{Julienne2014}. }
\end{figure*}

We consider three pairwisely interacting alkali-metal atoms of equal mass $m$ moving in an external magnetic field $\vb{B}$. The system is described using the Hamiltonian $H = H_0 + V$, where,
\begin{equation}
H_0 = \sum_{j} \left[T_j + H_j^{\mathrm{hf}}  + H_j^{\mathrm{Z}} (\vb{B})\right] , \qquad V = \sum_{\alpha} V_{\alpha}.
\label{eq:H}
\end{equation}
We use Latin indices $j$ to label the three particles and Greek indices $\alpha \equiv (j j')$ to label the three distinct pairs. The non-interacting Hamiltonian $H_0$ consists of the single-particle kinetic energy operators $T_j$ and the hyperfine and Zeeman interactions,
\begin{equation}
H_j^{\mathrm{hf}}  =  A_j^{\mathrm{hf}} \vb{s}_j \cdot \vb{i}_j, \qquad
H_j^{\mathrm{Z}}(\vb{B})  =   \left(\gamma_j^n \vb{i}_j + \gamma_j^e \vb{s}_j \right) \cdot \vb{B}.
\end{equation}
Here $\vb{s}_j$ and $\vb{i}_j$ respectfully represent the electronic and nuclear spin of atom $j$, $A_j^{\mathrm{hf}}$ is the hyperfine constant, and $\gamma_j^{n/e}$ are the electronic and nuclear gyromagnetic ratios. Together, the hyperfine and Zeeman Hamiltonians have 8 single-particle eigenstates, which can be unambiguously connected to distinct hyperfine states $\ket*{f_j, m_{f_j}}$ at zero magnetic field. In the three-body sector, spin-exchange couples all states with the same total $M_F = m_{f_1} + m_{f_2} + m_{f_3}$, resulting in the three-body spin spectra shown in Figs.~\ref{fig:3bodyMC}(a) and \ref{fig:3bodyMC}(b). Here the jump in complexity between the $\ket*{f, m_f}_{\mathrm{in}} = \ket*{1, 1}$ ($M_F = 3$) and $\ket*{f, m_f}_{\mathrm{in}} = \ket*{1, 0}$ ($M_F = 0$) calculations is immediately evident, and in addition one will note that the incoming state $\ket*{f, m_f}_{\mathrm{in}} = \ket*{1, 0}$ couples to several states that have nearly degenerate scattering thresholds \cite{Yudkin2023_abc}. In numerical practice we compose the three-body spin basis by forming products of two-body spin states with a third single-particle state, allowing us to explicitly (anti)symmetrize the two-body part in accordance with bosonic exchange symmetry \cite{Secker2021_2C}. This structure for the spin-basis also turns out to be the most well suited for the three-body equations, which we will discuss further down below.

Let us now define the two-body interactions $V_{\alpha}$, which we formulate using the method of Ref.~\cite{Julienne2014}. Neglecting magnetic dipole-dipole interactions, the potential is isotropic and can be decomposed as,
\begin{equation}
V_{\alpha} = V_{\alpha}^{0} \mathcal{P}_{\alpha}^0 + V_{\alpha}^1 \mathcal{P}_{\alpha}^1.
\end{equation}
where $\mathcal{P}_{\alpha}^0, \mathcal{P}_{\alpha}^1$ respectively project on the electronic singlet and triplet subspaces of the pair $\alpha$. At large two-body separations $r_{\alpha}$ the potentials are equivalent and given as $V_{\alpha}^{0/1}(r_{\alpha}) \rightarrow - C_6/r_{\alpha}^6$, where the dispersion coefficient $C_6$ defines the van der Waals length as $r_{\mathrm{vdW}} = (m C_6/\hbar^2)^{1/4}/2$. In the short-range, the shape of the singlet and triplet interactions is very complicated and depends strongly on non-universal detail of the atomic structure. We use the Born-Oppenheimer potentials formulated in Ref.~\cite{Julienne2014} by fitting two-body coupled-channels calculations to experimental scattering data and binding energies. The resulting potentials are plotted in Fig.~\ref{fig:3bodyMC}(c). 

\subsection{Three-body integral equations}
We now turn to the three-body problem. To evaluate the recombination rate we require the matrix elements of the three-body transition operator $U_{\alpha 0}(z)$, which we calculate from the Alt-Grassberger-Sandhas (AGS) equations \cite{Alt1967}. They can be most easily derived by starting from the two-body transition operator $t(\varepsilon)$ at two-body energy $\varepsilon$, which obeys the Lippmann-Schwinger equation,
\begin{equation}
t(\varepsilon) = V + V \frac{1}{\varepsilon - H_0^{\mathrm{2b}} - V} V
\label{eq:LS}
\end{equation}
Intuitively, $t(\varepsilon)$ describes how two particles coming in from infinity with energy $\varepsilon$ scatter with each other, and escape back to infinity. In the first order Born approximation, the second term is neglected, such that the whole process contains a single two-body interaction. This approximation fails at stronger interactions, where the multiple scatterings captured by the second term have to be included. 

To define an analogous transition operator for the three-body problem we have to account for the fact that now the interaction $V$ need not always vanish at infinity. This is immediately evident for the process of three-body recombination, where the system transitions from a free state (typically indicated with subscript 0) to an atom-dimer state (labelled with index $\alpha$ denoting the pair forming the molecule). With this in mind, the proper generalization of the Lippmann-Schwinger equation at three-body energy $E$ reads,
\begin{equation}
U_{\alpha 0}(E) = \sum_{\beta \neq \alpha} V_{\beta} + \sum_{\beta \neq \alpha} V_{\beta} \frac{1}{E - H_0 - V} V,
\end{equation}
where one notes that the interaction $V_{\alpha}$ between the molecular pair is excluded from the outgoing interaction vertex \cite{Glockle1983}. After some algebra, the right hand side may be written as,
\begin{equation}
U_{\alpha 0}(E) = \sum_{\beta \neq \alpha} \mathcal{T}_{\beta} + \sum_{\beta \neq \alpha} \mathcal{T}_{\beta} \frac{1}{E - H_0} U_{\beta 0},
\end{equation}
which is a form of the AGS equation. Here, 
\begin{equation}
\mathcal{T}_{\alpha}(E) = V_{\alpha} + V_{\alpha} \frac{1}{E - H_0 - V_{\alpha}} V_{\alpha},
\label{eq:Tau}
\end{equation}
is the \textit{off-shell} two-body transition matrix generalized to the three-body space. The fact that the three-body problem embeds the off-shell two-body problem logically follows from the third particles ability to carry away excess energy. Finally, we rewrite the AGS equation into symmetrized form by writing $U_{\alpha 0} \rightarrow \frac{1}{3} U_{\alpha 0} (1 + P)$, where $P = P_+ + P_-$ is the sum of cyclic permutation operators. Then we finally obtain the integral equation,
\begin{equation}
U_{\alpha 0}(E)  = \frac{1}{3} P \mathcal{T}_{\alpha}(E) (1 + P)  +  P \mathcal{T}_{\alpha}(E) G_0(E) U_{\alpha 0}(z). 
\label{eq:AGS}
\end{equation}
where we have also used the permutation operators to replace the sums over $\beta$, and defined $G_0(E) = (E-H_0)^{-1}$ as the uncoupled three-body Green's operator. Similar to Eq.~\eqref{eq:LS} the first term describes a single coupling between the pair $\alpha$ that forms the molecule and the third particle, which is captured to all orders of perturbation theory by $\mathcal{T}_{\alpha}(E)$. The second term expands into an infinite perturbation series that captures all multiple couplings, which become important for strong interactions.

We express Eq.~\eqref{eq:AGS} in the basis of partial-wave three-body states $\ket*{kq(l \lambda) L M_L} \otimes \ket*{ \overline{c_1 c_2}, c_3}$ \cite{Glockle1983}, where $k$ and $q$ are the magnitude of the relative dimer and atom-dimer momenta, $l$ and $\lambda$ are the relative dimer and atom-dimer angular momenta, and $L$ and $M_L$ are the total angular momentum and its magnetic projection. As the full system has rotational symmetry both $L$ and $M_L$ are conserved. We restrict our calculations to the lowest channel $L=0$, which dominates the three-body recombination rate for bosons in the low-energy limit \cite{Esry2001}. This immediately enforces that $l = \lambda$, such that the internal rotational states of the three-body complex can be described with a single quantum number. The two-body spin state $\ket*{\overline{c_1 c_2}}$ is (anti)symmetrized in accordance with the spatial parity set by $l$ \cite{Secker2021_2C}.

By the above definitions, we have a unique two-body transition operator $\mathcal{T}_{\alpha}^{c_3 l}(E)$ for each two-body spin projection $m_{f_{12}} = M_F - m_{f_3}$ and partial wave $l$. We obtain the associated two-body eigenenergies $\varepsilon_{\nu}^{c_3 l}$ and eigenfunctions $\ket*{\psi_{\nu}^{c_3 l}}$ by solving the two-body Schr\"odinger equation in position space using a mapped discrete variable representation \cite{Willner2004}. This approach constrains the movement of particles to a hard-wall box, whose size far exceeds all relevant length scales in the problem \cite{Secker2021_Grid}. With the two-body problem solved, we can formulate the complete set of eigenstates of the three-body Hamiltonian $H_0 + V_{\alpha}$ as $\ket*{\psi_{\nu}^{c_d l}} \otimes \ket*{q, c_3}$. Inserting into Eq.~\eqref{eq:Tau} we then have,
\begin{equation}
\mathcal{T}_{\alpha}^{c_3 l}(E) = V_{\alpha}  +   4 \pi \int_0^{\infty} dq \ q^2 \ket*{q, c_3} \sum_{\nu} \frac{V_{\alpha} \dyad*{\psi_{\nu}^{c_3 l}}{\psi_{\nu}^{c_3 l}} V_{\alpha}}{E - \varepsilon_{\nu}^{c_3 l} - \frac{3}{4}\frac{\hbar^2 q^2}{m} - \varepsilon_{c_3}} \bra*{q, c_3}, 
\label{eq:TauSpec}
\end{equation}
where $\varepsilon_{c_3}$ is the energy of the third particle spin state. In numerical practice we compute the spectral decomposition of this operator, which transforms the AGS equation \eqref{eq:AGS} to a one-dimensional integral equation that, upon discretization of the third particle momenta $q$, becomes a matrix equation that can be straightforwardly solved numerically. Our discretization, taken from Refs. \cite{Mestrom2019,Secker2021_Grid}, is based on Gauss-Legendre quadrature segmented into three sections as $q_1 r_{\mathrm{vdW}} \in \left[10^{-5}, 10^{-2} \right]$, $q_2 r_{\mathrm{vdW}} \in \left[10^{-2}, 1 \right]$ and $q_3 r_{\mathrm{vdW}} \in \left[1, q_{\mathrm{max}} r_{\mathrm{vdW}} \right]$. The value of $q_{\mathrm{max}}$ acts as an ultraviolet cut-off, which we discuss in more detail in Sec. \ref{sec:Num}. The number of grid points in each segment are tuned for convergence of $a_-$, which is typically very fast due to the efficiency of Gaussian quadrature. As is typical in quantum scattering theory we evaluate the AGS equation at infinitesimally complex energies, which allows us to converge the integration of the poles in the two-body at transition matrix at $q = \sqrt{4 m(E - \varepsilon_{c_3} - \varepsilon_{\nu}^{c_3 l}) /3\hbar^2}$, correspondent with the third-particle momenta where the three-body recombination process is exactly on-shell \cite{Mestrom2019}. 

\subsection{Three-body spin models}

To separate the contributions from two-body and three-body spin-exchange processes, we use two different models for the spin coupling \cite{Secker2021_39K}. In a full multichannel spin (FMS) model,
we make no approximations and include all three-body channels that conserve the total spin projection $M_F = m_{f_1} + m_{f_2} + m_{f_3}$. In the fixed spectating spin (FSS) model, we turn off three-body spin exchange processes by replacing the two-body interaction $V_{\alpha}$ with the restricted form,
\begin{equation}
V_{\alpha}^{\mathrm{FSS}} = V_{\alpha} \ket*{c_3^{\mathrm{in}}}\bra{c_3^{\mathrm{in}}}.
\end{equation}
In this way the two-body interaction is turned off if the third particle is not in the incoming spin state. In the AGS equations this replacement is equivalent to restricting $\ket*{c_3}$ to $\ket*{c_3^{\mathrm{in}}}$ everywhere, such that $m_{f_{12}} = m_{f_1} + m_{f_2}$ and $m_{f_3}$ are separately conserved.

\subsection{Fitting of the recombination rate}

For large negative scattering length, universal theory predicts a power-law increase of the three-body recombination rate coefficient, given as $L_3 = 3 C_{-}(a) \hbar a^4/m$. The function $C_{-}(a)$ gives the log-periodic oscillation of the recombination rate arising from the Efimov effect. A two-parameter analytical prediction for $C_-(a)$ can be obtained in the formalism of effective field theory, which gives \cite{Braaten2006},
\begin{equation}
C_-(a) = 4590 \ \frac{\sinh(2 \eta_-)}{\sin^2\left[s_0 \ln(a/a_-)\right] + \sinh^2(\eta_-)}.
\label{eq:Cm}
\end{equation}
To perform the fit of $L_3$ we first compute the dimensionless quantity $mL_3/(3\hbar a^4)$, and subsequently fit $C_-(a)$ directly. Provided that we compute $L_3$ close to an Efimov resonance, then 3 data points are typically sufficient to obtain a converged fit.

\subsection{Numerical convergence}
\label{sec:Num}

\begin{figure*}[t]
\subfloat{\includegraphics{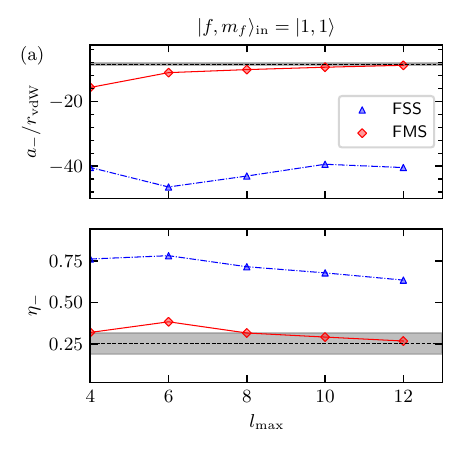}}
\subfloat{\includegraphics{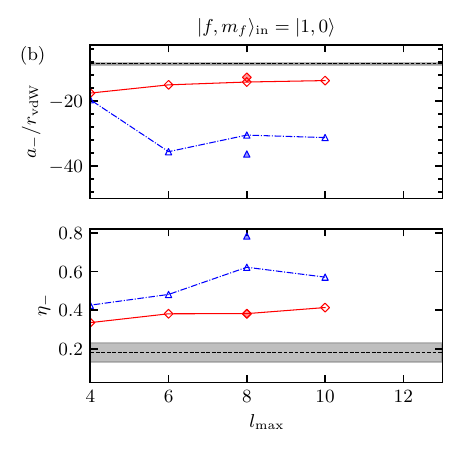}}
\caption{\label{fig:amin} Convergence of the three-body parameter $a_-$ and trimer width $\eta_-$ with the maximum two-body angular momentum $l_{\mathrm{max}}$, for the $\ket*{1,1}$ incoming hyperfine state in (a) and the $\ket*{1,0}$ incoming hyperfine state in (b). Filled scatter plots have $q_{\mathrm{max}} r_{\mathrm{vdW}} = 40$, while empty scatter plots have $q_{\mathrm{max}} r_{\mathrm{vdW}} = 20$. Colored lines have been added as a guide for the eye. The experimental data and uncertainty interval of Ref. \cite{Gross2011} is shown as a black dotted line with surrounding shaded region.}
\end{figure*}

Our numerical method is based on two important numerical parameters relevant for convergence, namely the maximum two-body partial wave $l_{\mathrm{max}}$ and the maximum third particle momentum $q_{\mathrm{max}}$ (which also limits the maximum binding energy for accessible molecular levels). In an FMS calculation, the number of coupled spin-channels is set purely by the initial state, in accordance with the conservation of the total magnetic quantum number $M_F$. As also noted in the main text, this means that an FMS calculation for $\ket*{f, m_f}_{\mathrm{in}} = \ket*{1,0}$, with a total of 50 coupled three-body channels, is considerably more expensive numerically than the $\ket*{f, m_f}_{\mathrm{in}} = \ket*{1,1}$ state, which has 22 coupled channels (for $l$ even). Hence, in the $\ket*{f, m_f}_{\mathrm{in}} = \ket*{1,1}$, we can fully converge an FMS calculation with parameters $l_{\mathrm{max}} = 12$ and $q_{ \mathrm{max}} r_{\mathrm{vdW}} = 40$, as we have confirmed by computing single points for $L_3$ with $l_{\mathrm{max}} = 16$ or $q_{\mathrm{max}} r_{\mathrm{vdW}} = 60$. For the $\ket*{f, m_f}_{\mathrm{in}} = \ket*{1,0}$ state, our numerical resources limit us to $l_{\mathrm{max}} = 8$ and $q_{\mathrm{max}} r_{\mathrm{vdW}} = 40$, which is not yet converged in $l_{\mathrm{max}}$. Typically, FMS calculations are limited by the considerable memory required to store the kernel of the STM equation. For example, the $l_{\mathrm{max}} = 8$ and $q_{\mathrm{max}} r_{\mathrm{vdW}} = 40$ FMS calculation in the $\ket*{f, m_f}_{\mathrm{in}} = \ket*{1,0}$ state required over 6 terabytes of internal memory, which was the biggest calculation we could perform. As the memory requirement will grow quadratically with $l_{\mathrm{max}}$, a converged calculation in this state will likely be considerably more taxing numerically.

However, despite the lack of numerical convergence, a comparison of the FMS results in Fig. \ref{fig:amin}(a) and Fig. \ref{fig:amin}(b) shows that for both states the value of $\abs{a_-}$ decreases monotonously with $l_{\mathrm{max}}$, although the convergence is slower for the $\ket*{f, m_f}_{\mathrm{in}} = \ket*{1,0}$ state likely due to the larger number of coupled channels. Extrapolating the trend for $a_-$ in Fig. \ref{fig:amin}(b) to larger values of $l_{\mathrm{max}}$ suggests that a converged result for the three-body parameter will further improve on the match with experimental data.

\section{Partial recombination rates}

\begin{figure*}[t]
\includegraphics{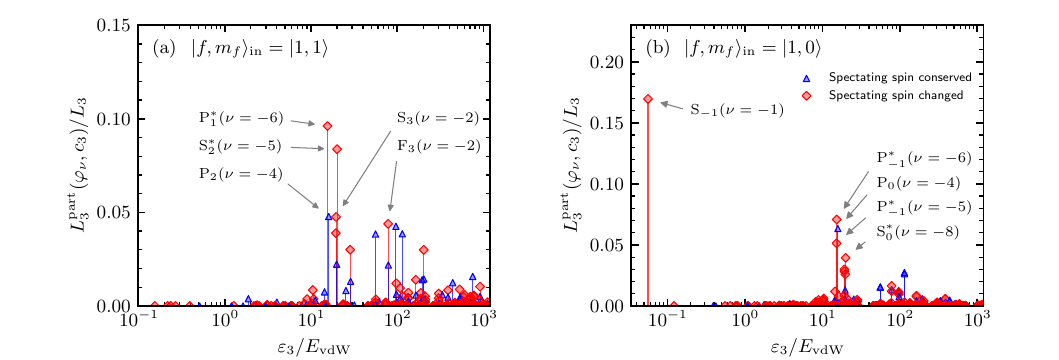}
\caption{Partial recombination rates $L_3^{\mathrm{part}}(\varphi_\nu, c_3)$ to individual atom-dimer channels as a function of the relative kinetic energy $\varepsilon_3 = 3\hbar^2 q^2/(4m)$ of the third particle following recombination, computed at a negative scattering length close to $a_-$. Again we distinguish between product channels accessible purely by two-body spin-exchange in blue, and only accessible by three-body spin-exchange in red. The 5 dominant product states are explicitly labelled using the notation $l_{m_{f_{12}}}(\nu)$, where $l$ is the molecular partial wave in spectroscopic notation, $m_{f_{12}} = m_{f_1} + m_{f_2}$ is the molecular magnetic quantum number and $\nu$ is the vibrational quantum number, defined such that $\nu = -1$ denotes the shallowest bound state in the associated channel. An asterisk denotes states in which the spectating particle is excited to the $f=2$ hyperfine manifold.}
\label{fig:PartRecomb}
\end{figure*}

To further highlight the significance of three-body spin-exchange in the system we analyze the partial recombination rates $L_3^{\mathrm{part}}(\varphi_\nu, c_3)$ explicitly. They are plotted in Fig.~\ref{fig:PartRecomb} as a function of the kinetic energy of the third particle following recombination.
Here it becomes evident that the majority of product states are weakly coupled, which gives the first hint for the existence of the propensity rule introduced in the main text. We also highlight the strongly dominant product state $S_{-1}(\nu = -1)$ for the incoming state $\ket*{f, m_f}_{\mathrm{in}} = \ket*{1,0}$, which is in fact a Feshbach dimer originating from a Feshbach resonance in the $m_{f_1} + m_{f_2} = -1$ two-body channel at $B = 938 \ \mathrm{G}$. Combined with the very small energetic separation from the incoming channel, see Fig.~\ref{fig:3bodyMC}(b), this makes that the kinetic energy following recombination is very small. This, combined with the enhanced scattering length in this channel leads to strongly enhanced three-body recombination. Note that the state $S_{-1}(\nu = -1)$ also conforms with the electronic spin-propensity rule as outlined in the main text. Earlier work has found the exact same state to provide an important recombination channel for this incoming hyperfine state in the weakly interacting regime \cite{Li2022}. Note that for the incoming state $\ket*{f, m_f}_{\mathrm{in}} = \ket*{1,1}$ no such special situation exists, and all closed-channel thresholds have much larger energetic spacing from the incoming state, see Fig.~\ref{fig:3bodyMC}(a).

\bibliography{References_sup}